\documentclass[aps,preprint]{revtex4}%
\usepackage{amsfonts}
\usepackage{amsmath}
\usepackage{amssymb}
\usepackage{graphicx}%
\begin{document}
\title[Short title for running header]{Spin Charge Recombination in Projected Wave Functions.}
\author{Hong-Yu Yang and Tao Li}
\affiliation{Center for Advanced Study, Tsinghua University, Beijing 100084, P.R.China}

\begin{abstract}
We find spin charge recombination is a generic feature of projected wave
functions. We find this effect is responsible for a series of differences
between mean field theory prediction and the result from projected wave
functions. We also find spin charge recombination plays an important role in
determining the dissipation of supercurrent, the quasiparticle properties and
the hole - hole correlation.

\end{abstract}
\volumeyear{year}
\volumenumber{number}
\issuenumber{number}
\eid{identifier}
\startpage{1}
\endpage{10}
\maketitle

Superconductivity results from Bose condensation of charged particles. In the
BCS theory of superconductivity, Fermionic electrons are paired into Bosonic
Copper pairs whose condensation lead to superconductivity. Soon after the
discovery of high temperature superconductors, Anderson proposed an exotic way
toward superconductivity in this class of materials. His way is to
fractionalize the electron rather than pair them up\cite{Anderson}%
\cite{Baskran}\cite{Zou}. The parent compounds of high temperature
superconductors are antiferromagnetic insulators. Anderson argued that doping
holes into such antiferromagnetic insulators would generate a spin liquid
state which can be envisioned as coherent superposition of spin singlet pairs.
He also argued that the excitations on the spin liquid state are
fractionalized. Specifically, the spin and charge quantum number of the
electron are now carried separately by two kinds of excitations, namely a
spin-$\frac{1}{2}$ chargeless Fermionic excitation called spinon and a
spinless Bosonic charged excitation call holon. In such a spin-charge
separated system, the charged holon is liberated from the Fermionic statistics
of the original electron and are ready to condense into a superfluid.

One main problem for such a proposal lies in the fact the predicted $T_{c}$ is
too high\cite{Kotliar}\cite{Lee1}. There is just no sufficient dissipation to
suppress the holon supercurrent. In the BCS theory, the supercurrent is
suppressed by quasiparticle excitation. These charged Fermionic excitation
form the normal fluid and cause dissipative response in external
electromagnetic(EM) field. However, in an ideal spin charge separated system,
the Fermionic spinon excitation dose not carry charge and do not cause
dissipation in an external EM field while the bosonic excitation of the holon
system is much less effective in dissipate the supercurrent.

The spin charge separation idea is nicely embodied in the slave Boson scheme
of $t-J$ model. In this scheme, the electron operator $c_{i\sigma}$ is written
as $f_{i\sigma}b_{i}^{\dagger},$ where $f_{i\sigma}$ and $b_{i}^{{}}$are
Fermionic spinon operator and Bosonic holon operator. Within this scheme, Lee
and Wen proposed that the spinon - holon recombination may hold the key for
the problem of overestimated $T_{c}$\cite{Lee2}. Through such a recombination,
the Fermionic spinon excitation acquire charge and can cause dissipation in EM
field, or, a charged hole regain Fermionic statistics and is transformed into
a \ normal carrier out of the condensate. However, it is not clear what is the
cause and nature of such a spinon - holon recombination. Wen and Lee argued
that the recombination may be related the unbroken $U(1)$ gauge structure in
their $SU(2)$ gauge theory of high temperature superconductivity\cite{Wen}.
The problem of spinon - holon recombination is also discussed
phenomenologically by Lee \textit{et. al.}\cite{Lee3} and Ng\cite{Ng}.

Here we point out that spinon - holon recombination is a generic feature in
projected wave functions. First we mention some clues that imply this. As our
first example, we consider the motion of holon in the so called uniform RVB
state on square lattice. The uniform RVB state is generated by the following
mean field ansatz%

\[
H_{f}=-%
{\displaystyle\sum\limits_{\left\langle ij\right\rangle ,\sigma}}
(f_{i,\sigma}^{\dagger}f_{j,\sigma}+h.c.)-\mu_{f}%
{\displaystyle\sum\limits_{i,\sigma}}
f_{i,\sigma}^{\dagger}f_{i,\sigma}%
\]

in which the sum is over nearest neighboring(NN) sites on the square lattice.
At the mean filed level, the motion of a holon in such a spin background is
described by the mean filed Hamiltonian%

\[
H_{h}=-t\chi%
{\displaystyle\sum\limits_{\left\langle ij\right\rangle }}
(b_{i}^{\dagger}b_{j}+h.c.)-\mu_{b}%
{\displaystyle\sum\limits_{i}}
b_{i}^{\dagger}b_{i}%
\]

in which\ $\chi$ is the mean filed hopping matrix element in such a spin
background and is given by $\chi=%
{\displaystyle\sum\limits_{\sigma}}
\left\langle f_{i,\sigma}^{\dagger}f_{j,\sigma}\right\rangle ,$ in which $i$
and $j$ are NN sites. At the mean filed level, the ground state of the system
is given by%

\[
\left\vert FS\right\rangle =(b_{q=0}^{\dagger})^{N_{b}}%
{\displaystyle\prod\limits_{k<k_{F}}}
f_{k\uparrow}^{\dagger}f_{-k\downarrow}^{\dagger}\left\vert 0\right\rangle
\]
where $N_{b}$ is the number of holon , $k_{F}$ is the spinon Fermi
surface(FS). In the mean filed ground state, $\chi=%
{\displaystyle\sum\limits_{k<k_{F}}}
(\cos(k_{x})+\cos(k_{y}))$ and is nonzero. Thus each holon has a kinetic
energy of order $t\chi$. Now we project the mean field ground state into the
physical subspace of no double occupancy. The projection of the spinon wave
function lead to the uniform RVB state(the projection of the holon condensate
only contribute a constant)%

\[
\left\vert U-RVB\right\rangle =P_{G}\left\vert FS\right\rangle =P_{G}\left(
{\displaystyle\sum\limits_{i,j}}
a_{_{ij}}f_{i\uparrow}^{\dagger}f_{j\downarrow}^{\dagger}\right)
^{\frac{N_{f}}{2}}\left\vert 0\right\rangle
\]

with the RVB amplitude $a_{ij}$ given by $%
{\displaystyle\sum\limits_{k<k_{F}}}
e^{ik(i-j)}$. At half filling, the RVB amplitude $a_{ij\text{ }}$has the
important characteristics that it is nonzero only for sites $i$ and $j$
belonging to different sublattices. Thus, contrary to our expectation from
mean filed theory, the holon in fact can not hop between NN sites in such a
spin background. One would argue that the spin wave function should be
modified upon hole doping. According to mean filed theory, the most natural
guess on the modification is to remove $N_{b}$ spinon below the spinon Fermi
surface. For two holes, the modified mean filed state is given by%

\[
\left\vert MF^{\prime}\right\rangle =(b_{q=0}^{\dagger})^{2}f_{k_{0}\uparrow
}^{{}}f_{-k_{0}\downarrow}^{{}}%
{\displaystyle\prod\limits_{k<k_{F}}}
f_{k\uparrow}^{\dagger}f_{-k\downarrow}^{\dagger}\left\vert 0\right\rangle
=f_{k_{0}\uparrow}^{{}}f_{-k_{0}\downarrow}^{{}}\left\vert FS\right\rangle
\]
in which $k_{0}$ and $-k_{0}$ are momentums below the spinon Fermi surface
where a pair of spinons are removed. This wave function represents a state
with two holons at $q=0$ and two spinon excitations(more exactly, two holes of
spinon) at $k_{0}$ and $-k_{0}$ on the half filled uniform RVB background.
Projecting this mean field state into the subspace of no double occupancy, we
get a RVB state with a modified RVB amplitude $a_{ij}^{\prime}=\frac{1}{N}%
{\displaystyle\sum\limits_{k<k_{F},k\neq k_{0}}}
e^{ik(i-j)}$. The change of the RVB\ amplitude caused by the spinon excitation
is vanishingly small(of order $1/N$, where $N$ is number of lattice sites) and
it seems that the hole motion between NN sites is still blocked. However, by
direct calculation of kinetic energy in the modified RVB\ state, we find such
an expectation is wrong. The kinetic energy per hole is of order $t$ rather
than vanishingly small. The only explanation for this surprising result is
that the spinon excitation is bound to the moving holon. If the spinon
excitation and the holon are independent of each other, the change of the
local spin background around the moving holon caused by the spinon excitation
would be of order $1/N$ and would not be able to release to NN kinetic energy.

The uniform RVB state is quite special. However, the spinon - holon
recombination is quite generic in projected wave functions. Now we consider
the d-wave RVB\ state on the square lattice generated by the ansatz%

\[
H_{f}=-%
{\displaystyle\sum\limits_{\left\langle ij\right\rangle ,\sigma}}
(f_{i,\sigma}^{\dagger}f_{j,\sigma}+h.c.)+%
{\displaystyle\sum\limits_{\left\langle ij\right\rangle }}
\Delta_{ij}(f_{i,\uparrow}^{\dagger}f_{j,\downarrow}^{\dagger}+f_{j,\uparrow
}^{\dagger}f_{i,\downarrow}^{\dagger}+h.c.)-\mu_{f}%
{\displaystyle\sum\limits_{i,\sigma}}
f_{i,\sigma}^{\dagger}f_{i,\sigma}%
\]
in which $\Delta_{ij}=\Delta$ and $-\Delta$ for NN sites along $x$ and $y$
directions. The mean field ground state is given by%

\[
\left\vert d-BCS\right\rangle =(b_{q=0}^{\dagger})^{N_{b}}%
{\displaystyle\prod\limits_{k}}
(1+\frac{\Delta_{k}}{\xi_{k}+\sqrt{\xi_{k}^{2}+\Delta_{k}^{2}}}f_{k\uparrow
}^{\dagger}f_{-k\downarrow}^{\dagger})\left\vert 0\right\rangle
\]
in which $\xi_{k}$ and $\Delta_{k}{}$ are mean field kinetic energy and
pairing gap of the spinon. Projecting $\left\vert d-BCS\right\rangle $ into
the subspace of no double occupancy generates a RVB state with $a_{ij}=%
{\displaystyle\sum\limits_{k}}
\frac{\Delta_{k}}{\xi_{k}+\sqrt{\xi_{k}^{2}+\Delta_{k}^{2}}}e^{ik(i-j)}$. In
the d-wave RVB state, the NN hopping is not suppressed. However, the matrix
element for next nearest neighboring(NNN) and next next nearest
neighboring(NNNN) hopping is very small near half filling. This is reasonable
since the mean field matrix element for hoping between sites on the same
sublattice is exactly zero when $\mu_{f}=0$. Lee \textit{et. al. }find the
kinetic energy due to NNN and NNNN hopping can be released by creating spinon
excitation at appropriate momentums on the d- wave RVB state\cite{TKLee}%
\cite{note1}\cite{note2}. For the case of two holes, they find the NNN and
NNNN kinetic energy is released in a state with $a_{ij}^{\prime}=%
{\displaystyle\sum\limits_{k\neq k_{0}}}
\frac{\Delta_{k}}{\xi_{k}+\sqrt{\xi_{k}^{2}+\Delta_{k}^{2}}}e^{ik(i-j)}$. It
is easy to check that this state can be generated by projecting $\left\vert
MF^{\prime}\right\rangle =$ $f_{k_{0}\uparrow}^{{}}f_{-k_{0}\downarrow}^{{}%
}\left\vert d-BCS\right\rangle $ and thus represent a state with two spinon
excitations. Thus once again we see the creation of an individual spinon
excitation can make an order of one change on the hopping matrix element of
holon. This again indicate that the spinon excitation is bound to the moving holon.

\bigskip The spinon - holon recombination can be inferred also from the
quasiparticle weight. In the slave Boson mean field theory, the quasiparticle
weight is proportional to the holon condensate and vanish with hole density.
After projection, the quasiparticle weight can have a nonzero value even at
vanishingly small hole density. As a trivial example in this respect, we
consider doping a hole into a fully polarized spin background. Since the spin
is fully polarized, the system is in fact in a free particle state and the
quasiparticle weight should be exactly one. As we will show below, the
difference between the prediction from the mean filed theory and that from the
projected wave function can be understood as a result of spinon - holon
recombination. In fact, the spinon excitation and holon are totally combined
in the fully polarized spin background in the sense that they sit at the same
site and bind into a real electron. A a less trivial example we consider
doping a hole into a spin background with antiferromagnetic long range order.
As we will show below, the spinon excitation and the holon will form well
defined bound state in such a spin background. This bound state has a nonzero
overlap with a bare electron. Thus the quasiparticle weight do not vanish near
half filling in this case.

Now we define the spinon - holon recombination more concretely. For
simplicity, we consider the uniform RVB\ state and dope only one hole into the
system. The mean field wave function for the doped system is%

\[
\left\vert k_{0},\uparrow\right\rangle =b_{q=0}^{\dagger}f_{-k_{0}\downarrow
}^{{}}%
{\displaystyle\prod\limits_{k<k_{F}}}
f_{k\uparrow}^{\dagger}f_{-k\downarrow}^{\dagger}\left\vert 0\right\rangle
=\frac{1}{N}%
{\displaystyle\sum\limits_{i,j}}
e^{ik_{0}(i-j)}b_{i}^{\dagger}f_{j\downarrow}^{{}}%
{\displaystyle\prod\limits_{k<k_{F}}}
f_{k\uparrow}^{\dagger}f_{-k\downarrow}^{\dagger}\left\vert 0\right\rangle
\]
This wave function represents a state with a holon at $q=0$ (created by
$b_{q=0}^{\dagger})$ and a spinon excitation at $-k_{0}$ (created by
$f_{-k_{0}\downarrow}^{{}}$) on the half filled RVB\ background. In this mean
field state, the spinon excitation and the holon are independent of each
other. Now we discuss how they are correlated in the projected wave function.
Here we define the site on which $f_{j\downarrow}^{{}}$ operate as the
location of the spinon excitation and study how the spinon excitation is
distributed when the holon is located on site $i$. In fact, it suffers from
some ambiguity to talk about the location of the spinon excitation on the
projected wave function, especially when the RVB amplitude is long
ranged\cite{Read}. Although suffers from such ambiguity, the correlation
function defined above is still of great value for understanding the
difference between mean field theory and projected wave functions. For
example, if the correlation function reduce to a delta function, then the
spinon operator $f_{j\downarrow}^{{}}$ and the holon operator $b_{i}^{\dagger
}$ act on the same site and as a whole is equivalent to the operation of a
bare electron operator $c_{i\downarrow}$. In this case, the spinon and the
holon are recombined into a real electron.

The desired correlation function can be evaluated easily. Suppose the holon
sit on site $i$ while the spinon sit on a different site $j$. Since all sites
besides $i$ are singly occupied after the projection, site $j$ must be doubly
occupied before the action of $f_{j\downarrow}^{{}}$ while site $i$ must be
empty. Thus the probability for such a spinon - holon configuration is given
by the probability of finding site $i$ empty and site $j$ doubly occupied and
with all other sites singly occupied in the mean field state $\left\vert
FS\right\rangle $. At the same time, the probability for the holon and the
spinon to sit on the same site is given by the probability of finding site $i$
occupied by a down spin and with all other sites singly occupied in
$\left\vert FS\right\rangle $. The ratio between the two probability $P_{ij}$
and $P_{ii}$ is given by%

\[
\frac{P_{ij}}{P_{ii}}=\frac{%
{\displaystyle\sum\limits_{\beta}}
\left\vert \psi_{\beta}\right\vert ^{2}}{%
{\displaystyle\sum\limits_{\alpha}}
\left\vert \psi_{\alpha}\right\vert ^{2}}=\frac{%
{\displaystyle\sum\limits_{\alpha}}
\left\vert \psi_{\alpha}\right\vert ^{2}\frac{\left\vert \psi_{\beta
}\right\vert ^{2}}{\left\vert \psi_{\alpha}\right\vert ^{2}}}{%
{\displaystyle\sum\limits_{\alpha}}
\left\vert \psi_{\alpha}\right\vert ^{2}}%
\]
Here, $\alpha$ denotes an arbitrary configuration with all sites singly
occupied, $\beta$ denotes an arbitrary configuration with site $i$ empty and
site $j$ doubly occupied and with all other sites singly occupied. In deriving
this formula, we have used the fact that each configuration $\beta$ can be
generated from two configuration $\alpha$ through electron hopping from site
$i$ to site $j$. This statistical sum can be evaluated easily with Variational
Monte Carlo method.

\bigskip Now we present the result for the spinon - holon correlation function
in various projected wave functions. Figure 1 shows the correlation function
for the projected one dimension Fermi sea. The projected one dimensional Fermi
sea is found to be a very good variational guess on the ground state of one
dimensional $t-J$ model\cite{Yokoyama}. We see the spinon - holon correlation
function decay as $1/r$ at large distance in this state. This power law decay
(which is not integrable) lead to a vanishingly small quasiparticle weight
near half filling. Figure 2 show the result for the two dimensional d-wave RVB
state. The correlation function in this case also decay with power law at
large distance and is not integrable. Thus the quasiparticle weight in this
case also vanishes near half filling. However, numerically the power law
decaying tail is quite small for both the one dimension projected Fermi sea
and the d-wave RVB state. The power law tail is hardly visible in Figure 2 due
to its numerical smallness. In Figure 3 we plot the the dependence of
correlation at the largest distance of the lattice as a function of the
lattice size. From this plot we see the spinon - holon correlation decay
approximately as $1/r^{3/2}$ at large distance in the d-wave RVB state. 

Now we show some examples with more tightly bound spinon - holon pairs. The
first example is the fully polarized state. In this state, the probability of
finding a doubly occupied site is zero. Thus the spinon and the holon must
occupy the same site and recombine into a bare electron. Hence the
quasiparticle weight is exactly one. As our second example, we consider states
with antiferromagnetic long range order. In this case, the configuration with
site $i$ empty and site $j$ doubly occupied is separated from the
configuration with all sites singly occupied in energy by a gap proportional
to the SDW order parameter. Thus, we expect the spinon - holon correlation
function to decay exponentially at large distance. Our calculation do find
such an exponential decay as shown in Figure 4 and 5. This exponential decay
indicates that the spinon and the holon form well defined bound state and has
a finite overlap with a bare electron. Calculation of Lee \textit{et. al.} do
find a finite quasiparticle weight on such a state. \ \ \ \ 

\qquad Another consequence of the spinon - holon recombination is the change
of statistics of the charge carrier. In the absence of the spinon excitation,
the holon is a bosonic excitation which move coherently in the RVB background.
In the presence of the spinon excitation, the holon tend to bind with the
spinon. The composite object of spinon - holon pair then acquire Fermi
statistics and become normal carrier. This is especially true when the spinon
- holon bound state is well defined. In the case of power law decaying spinon
- holon correlation, there is no well defined bound state and the statistics
is in a strict sense not defined. However, when the energy scale involved is
not too small, assigning Fermi statistics to the composite object of spinon -
holon pair is reasonable since the power law decaying tail is numerically very small.

The spinon - holon recombination and the related change of statistics of
charge carrier is essential for the dissipation of the supercurrent in a spin
charge separated superconductor\cite{Lee2}. The thermally excited spinon
excitation would combine with the holon in the holon condensate. This
combination would transform a superconducting charge carrier into a normal
charge carrier. When the number of thermally excited spinon equals to the
number of holon, all charge carrier in the superfluid are transformed into
normal carrier and the superconductivity is gone. At low doping, the number of
spinon excitation needed to destroy the superconductivity is small and it is
thus expected that the spin state above $T_{c}$ is not significantly different
from the RVB ground state. This may explain the normal state spin gap observed
in underdoped cuprates.

As mentioned above, the spinon excitation can also be spontaneously generated
in the d- wave RVB state by nonbipartite(for example NNN and NNNN) hopping
term in the Hamiltonian. The spontaneously generated spinon will combine with
the holon in the condensate and transform the latter into normal carrier.
Thus, superconductivity is destroyed by the nonbipartite hopping term at low
doping. This effect is recently studied by Shih \textit{et. al.}\cite{Shih} At
higher doping level, the RVB background is modified so that the NNN and NNNN
hoping are not suppressed and there is no need to generate spinon excitation.
Then superconductivity will survive.

The spinon - holon recombination also modify the hole - hole correlation. When
the spinon and the holon are tightly bound, the composite object of spinon -
holon pair tend to avoid each other due to its Fermionic statistics. This
change of hole - hole correlation is observed in numerical work of Lee $et.$
$al$\cite{TKLee}$.$ In their work, they calculated the correlation of a pair
of holes in a state with coexisting d-wave RVB and SDW order. They find the
holes tend to attract each other when there is no spinon excitation. When a
pair of spinon excitations are generated, the hole - hole attraction
disappear. Figure 6and 7 show the hole - hole correlation calculated in the
d-wave RVB state with and without spinon excitation. Although spinon and holon
are less tightly bound in the d-RVB state, the influence of spinon - holon
recombination on the hole - hole correlation function is still quite
remarkable. From the figure we see clearly that the second hole is pushed away
from the hole at the origin in the presence of a pair of spinon excitations.

We conclude that spinon - holon recombination is a generic feature in
projected wave functions. This effect play an important role in the
dissipation of supercurrent in cuprates. The spinon - holon recombination also
affect significantly the quasiparticle properties and hole - hole correlation
in projected wave functions.

\bigskip This work is supported by NSFC Grant No.90303009. The authors would
like to thank members of the HTS group at CASTU for discussion.

%

\begin{figure}
[ptb]
\begin{center}
\includegraphics[
natheight=3.260300in,
natwidth=4.063900in,
height=2.5753in,
width=3.2021in
]%
{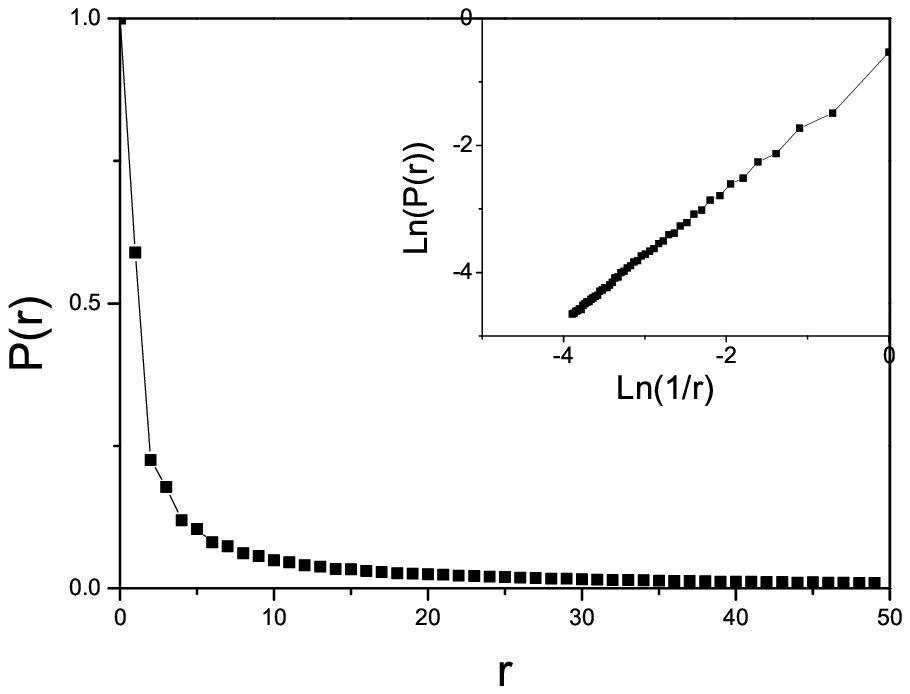}%
\caption{Spinon - holon correlation function for projected one dimensional
Fermi sea. The inset show the data in logarithmic scale.}%
\label{fig1}%
\end{center}
\end{figure}
\begin{figure}
[ptbptb]
\begin{center}
\includegraphics[
natheight=3.418800in,
natwidth=4.117000in,
height=2.5753in,
width=3.0942in
]%
{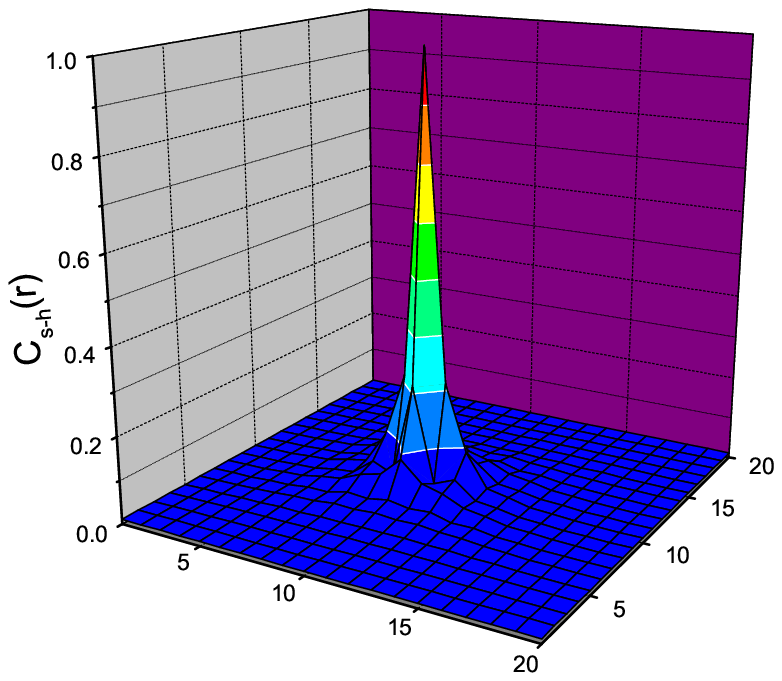}%
\caption{Spinon - holon correlation function for a d-wave RVB state with
$\Delta=0.25.$ The calculation is done on a $20\times20$ lattice and the holon
is located at (10,10). }%
\label{fig2}%
\end{center}
\end{figure}
\begin{figure}
[ptbptbptb]
\begin{center}
\includegraphics[
natheight=5.192200in,
natwidth=6.972200in,
height=2.5753in,
width=3.4487in
]%
{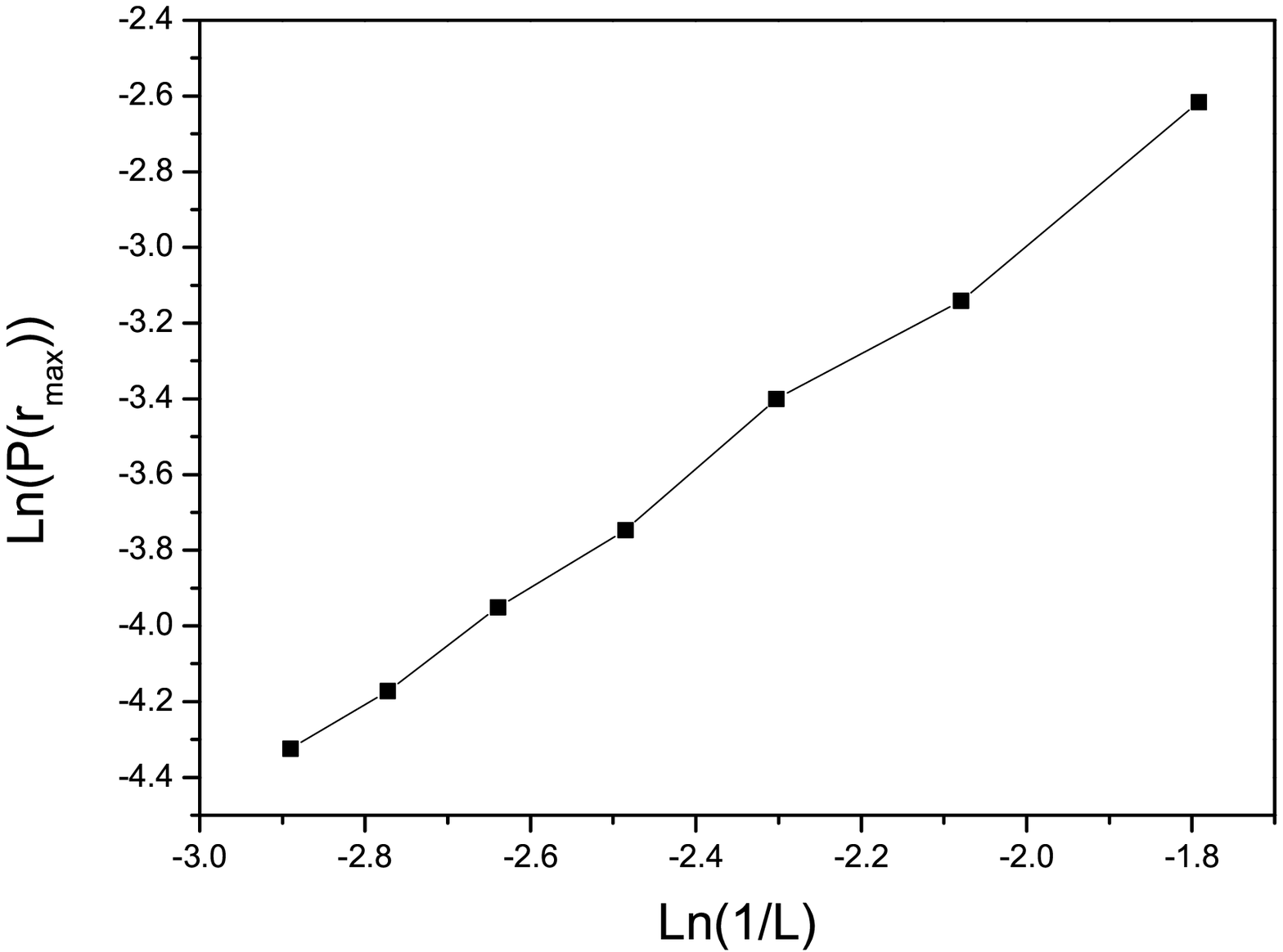}%
\caption{Power law decay of the spinon - holon correlation at large distance
for d-wave RVB state. Here, L is the lattice size, $r_{\max}$ is largest
distance that can be defined in such a lattice.}%
\label{fig2b}%
\end{center}
\end{figure}
\begin{figure}
[ptbptbptbptb]
\begin{center}
\includegraphics[
natheight=3.418800in,
natwidth=4.117000in,
height=2.5753in,
width=3.0942in
]%
{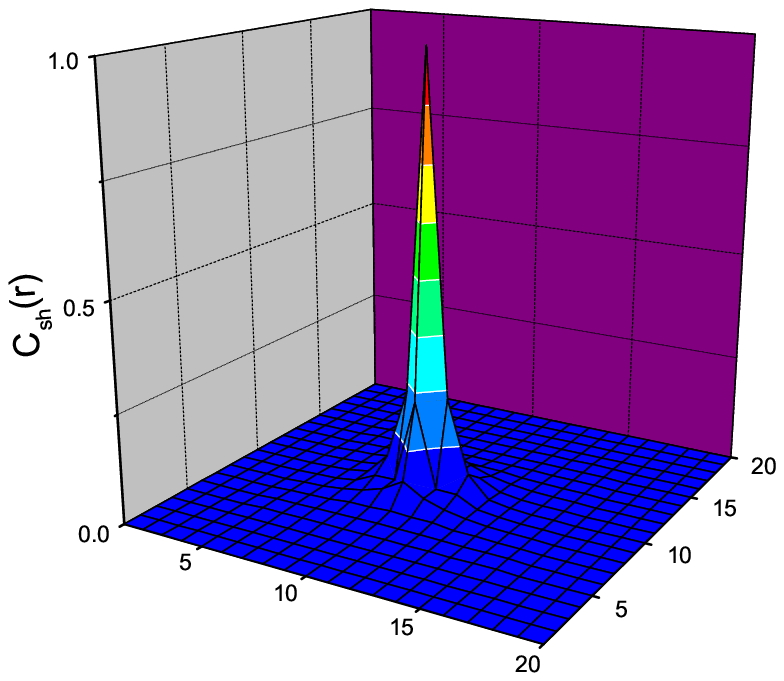}%
\caption{Spinon - holon correlation function in a spin background with both
d-wave RVB and antiferromagnetic order. The SDW order parameter is
$\Delta_{AF}=0.1.\Delta=0.25.$}%
\label{fig3}%
\end{center}
\end{figure}
\begin{figure}
[ptbptbptbptbptb]
\begin{center}
\includegraphics[
natheight=3.218800in,
natwidth=4.104600in,
height=2.577in,
width=3.2777in
]%
{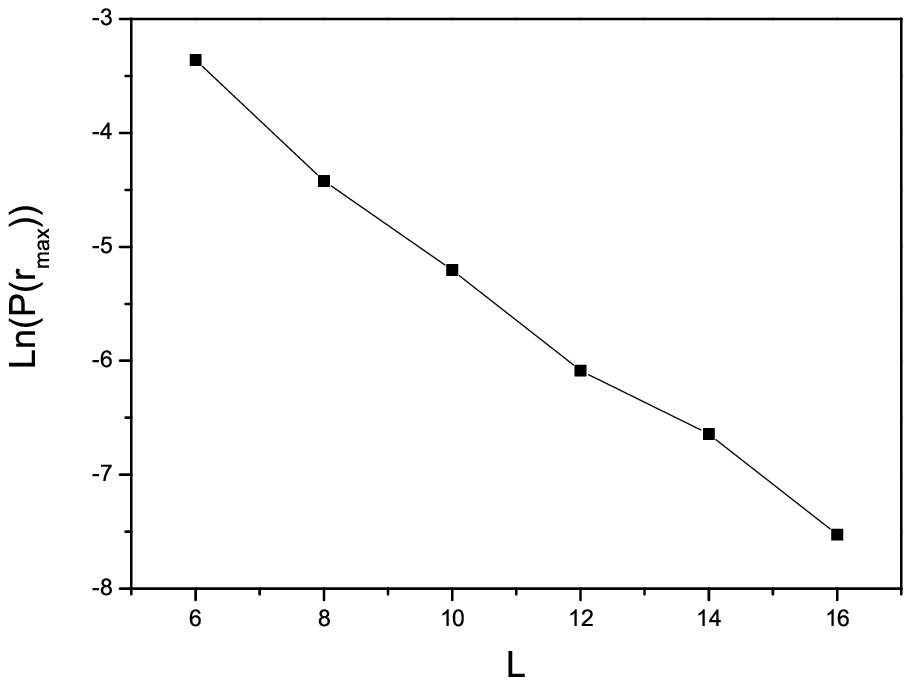}%
\caption{Exponential decay of spinon - holon correlation at large distance in
a spin background with antiferromagnetic order.}%
\label{fig3b}%
\end{center}
\end{figure}
\begin{figure}
[ptbptbptbptbptbptb]
\begin{center}
\includegraphics[
natheight=3.418800in,
natwidth=4.117000in,
height=2.5753in,
width=3.0942in
]%
{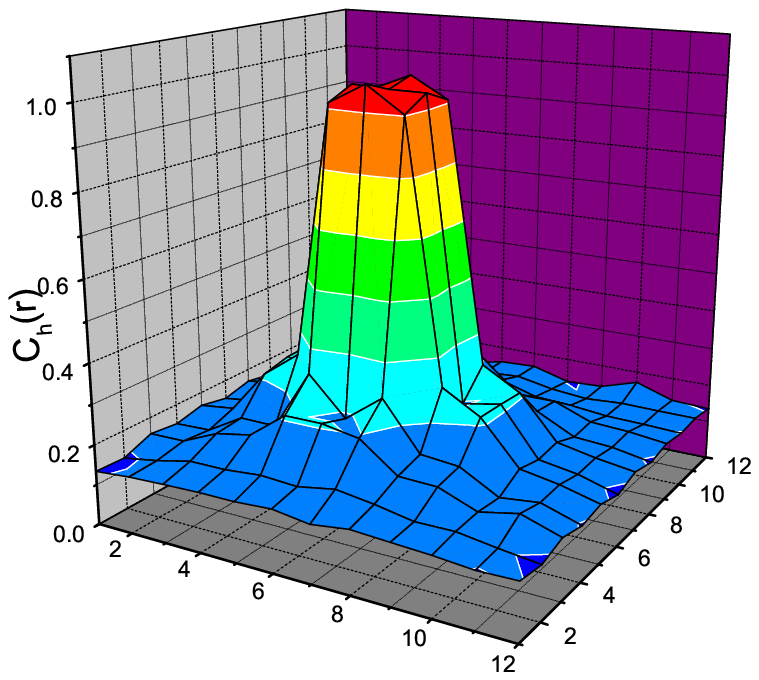}%
\caption{Hole - hole correlation (normalized by its value for NN holes) in a
d-wave RVB state with $\Delta=0.25$ in the absence of spinon excitation. The
calculation is done on $12\times12$ lattice and one the hole is located at
(6,6).}%
\label{fig4a}%
\end{center}
\end{figure}
\begin{figure}
[ptbptbptbptbptbptbptb]
\begin{center}
\includegraphics[
natheight=3.418800in,
natwidth=4.117000in,
height=2.5753in,
width=3.0942in
]%
{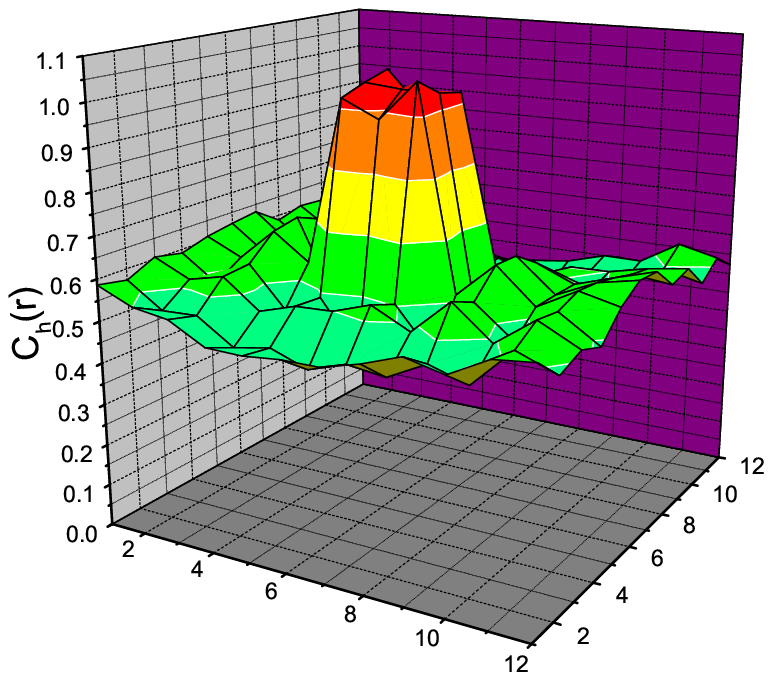}%
\caption{Hole - hole correlation in a d-wave RVB state with $\Delta=0.25$ in
the presence of a pair of spinon excitation.}%
\label{fig4b}%
\end{center}
\end{figure}

\medskip

\begin{thebibliography}{99}
\bibitem {note1}The state studied in \cite{TKLee} also has antiferromagnetic
long range order. This order enhance the spinon - holon recombination.
\bibitem {note2}Another way to release the NNN and NNNN kinetic energy is to
introduce NNN and NNNN hoping term in the mean field ansatz for the spinon.
However, such nonbipartite term will cause fundmental change on the structure
of spin wave function and result in significant increase of the exchange energy.
\end{thebibliography}
\end{document}